\documentclass{aa}
\usepackage{graphicx,natbib,url,txfonts,color}
\begin{document}
\title{First nonlinear force-free field extrapolations of SOLIS/VSM data}
\author{%
      J.~K.~Thalmann \inst{1},~ T.~Wiegelmann \inst{1}
      \and
      N.-E.~Raouafi\inst{2}%
}
\institute{%
      Max-Planck-Institut f\"ur Sonnensystemforschung, 37191 Katlenburg-Lindau, Germany\\
      \email{[thalmann;wiegelmann]@mps.mpg.de}
      \and
      National Solar Observatory, 85719 Tucson, Arizona\\
      \email{raouafi@noao.edu}
}
\date{\today}
\abstract
{}
{
We study the coronal magnetic field structure inside active regions and its temporal evolution. We attempt to
compare the magnetic configuration of an active region in a very quiet period with that for the same region
during a flare.
}
{
Probably for the first time, we use vector magnetograph data from the Synoptic Optical Long-term
Investigations of the Sun survey (SOLIS) to model the coronal magnetic field as a sequence of nonlinear
force-free equilibria. We study the active region NOAA 10960 observed on 2007 June 7 with three snapshots
taken during a small C1.0 flare of time cadence 10 minutes and six snapshots during a quiet period.
}
{
The total magnetic energy in the active region was approximately $3 \times 10^{25}$ J. Before the flare the
free magnetic energy was about 5~\% of the potential field energy. A part of this excess  energy was released
during the flare, producing almost a potential configuration at the beginning of the quiet period.
}
{
During the investigated period, the coronal magnetic energy was only a few percent higher than that of the
potential field and consequently only a small C1.0 flare occurred. This was compared with an earlier
investigated active region 10540, where the free magnetic energy was about 60~\% higher than that of the
potential field producing two M-class flares. However, the free magnetic energy accumulates before and is
released during the flare which appears to be the case for both large and small flares.
}
\keywords{Sun: magnetic fields -- Sun: flares -- Sun: corona}
\authorrunning{Thalmann et al.}
\titlerunning{First nonlinear force-free field extrapolations using SOLIS/VSM data}
\maketitle

\section{Introduction}

Methods have been developed to extrapolate the observed photospheric magnetic field vector into the
corona. Using the fact that the magnetic field is dominant in solar active regions (ARs), we are able
to neglect non-magnetic forces and to assume that the coronal magnetic field is force-free.
Different instruments provide photospheric vector magnetograph data, which are used as input to the
extrapolation methods. These data have, however had a rather low time cadence. Data of high time
cadence are required to investigate in detail, for example the different evolutionary stages of solar
flares. Suitable data include magnetic field observations of the Sun provided by the SOLIS
Vector-SpectroMagnetograph. With a time cadence of $\approx$ 10 minutes, the instrument is designed to
measure multiple area scans of ARs, which enables us for the first time to investigate the evolution of the
coronal magnetic field energy with a high time cadence. Many existing studies deal with the extrapolation
based on vector magnetograph data. For instance, \cite{reg_07b} dealt with the photospheric vector
magnetic field provided by the Mees Solar Observatory Imaging Vector Magnetograph, \cite{wie_05} used
spectropolarimetric data recorded with the Tenerife Infrared Polarimeter of the German Vacuum Tower
Telescope, and \cite{tha_08} performed extrapolations of Solar Flare Telescope Vector Magnetograph
data. In all of these studies, only one snapshot was however used or, as in the last of the aforementioned
studies, a sequence of vector magnetograms with a low time cadence of one magnetogram per day.
Therefore, an improvement is achieved by applying our extrapolation technique to the high time cadence
SOLIS/VSM data as described in the present study.
\section{Method}
\subsection{Instrumentation: The SOLIS/VSM instrument}
    \begin{figure}[ht]\centering
    \includegraphics[width=0.4\textwidth]{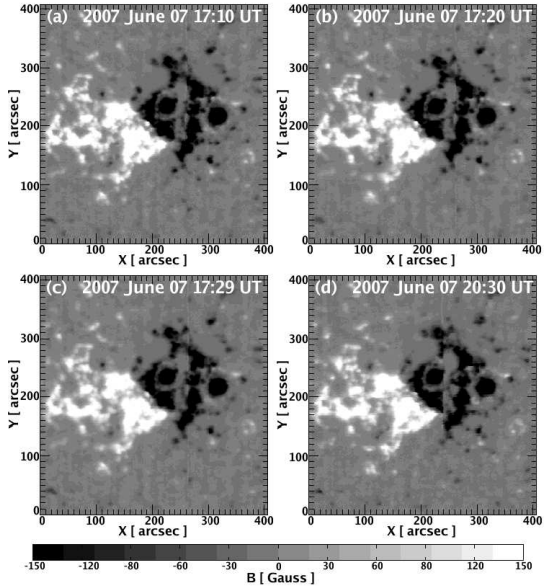}
    \caption{Longitudinal component of the SOLIS/VSM data during the C1.0 flare in panels (a) -- (c) and
    after the flare in panel (d).}
    \label{fig:fig1}
    \end{figure}
The Vector-SpectroMagnetograph \citep[VSM; see][]{jon_02} on the Synoptic Optical Long-term Investigations
of the Sun \citep[SOLIS; see][]{kel_har_03} has provided magnetic field observations of the Sun almost
continuously since August 2003. The instrument is designed to measure the magnetic field vector everywhere
on the solar disk. Full disk vector observations are completed at least weekly. Multiple areas scans of ARs
have also been available since November 2006. In addition, longitudinal magnetic field measurements in
the photosphere (at the Fe~{\sc{i}} 630.15 nm and 630.25 nm spectral lines) and chromosphere (at the
Ca~{\sc{ii}} 854.2 nm spectral line) are available on a daily basis. Quick-look data (JPEG images and FITS
files) of the magnetic field vector in and around automatically selected ARs \citep[][]{geo_08} are available
online to the community. The Quick-look data provides estimates of the magnetic field strength, inclination,
and azimuth \citep[][and references therein]{hen_06}, which should, because of the high field strength in ARs,
be comparable to fully inverted data only differing by a few percent. The azimuth 180$^\circ$-ambiguity is
solved using the Nonpotential Magnetic Field Calculation method \citep[NPFC; see][]{geo_05}, which does
not introduce any error in the azimuth or any other quantities. Tools for full Milne-Eddington inversion are
being developed to provide more accurate magnetic data especially in weak field regions.\\
In the time period around an C1.0 flare on 2007 June 7, three SOLIS/VSM vector magnetograms were available
to use. One in the rising phase of the emission, one at the time when the flare peaked, and one in its decaying
phase (at 17:10~UT, 17:20~UT, and 17:29~UT as shown in panel (a), (b), and (c) in Fig.~\ref{fig:fig1},
respectively). All the other magnetic field measurements on 2007 June 7 (between 20:30~UT and  21:42~UT)
allow us to investigate the magnetic field structure in a period of low solar activity.

\subsection{Numerics: Nonlinear force-free extrapolation}
The basic equations for the computation of the nonlinear force-free magnetic field vector $\mathbf{B}$ are
    \begin{equation}
        (\nabla \times \mathbf{B}) \times \mathbf{B} = 0, \label{equ:nolf}
    \end{equation}
    \begin{equation}
        \nabla \cdot \mathbf{B} = 0, \label{equ:sol}
    \end{equation}
where Eq.~(\ref{equ:nolf}) expresses that the Lorentz force is forced to vanish (as a consequence of \textbf{j}
$\parallel$ \textbf{B}, where \textbf{j} is the electric current density) and Eq.~(\ref{equ:sol}) describes the
absence of magnetic monopoles. For reviews on how to solve these equations, we refer the reader for example
to \cite{sak_89}, \cite{ama_97}, and \cite{wie_08}.\\
A special form of the force-free fields are potential magnetic fields which can be computed from the longitudinal
photospheric magnetic field alone and correspond to the minimum energy state for given boundary conditions.
We calculate the potential field with the help of a Fast-Fourier method \citep{ali_81}. In an AR, only the energy
exceeding that of a potential field -- the so-called free magnetic energy -- can partly be transformed into kinetic
energy during dynamic events. Therefore, nonlinear force-free (NLFF) field models are required for a realistic
estimation of the coronal magnetic field. Some existing methods for computing NLFF fields were tested and
compared by \cite{sch_06}, \cite{met_08}, and \cite{sch_08}. These works revealed that the optimization
method, proposed by \cite{whe_00} and implemented by \cite{wie_04}, was a reliable and fast algorithm. This
approach evolves the magnetic field to reproduce the boundary, force-free, and divergence-free conditions by
minimizing a volume integral of the form
      \begin{equation}
      L = \int_V w(x,y,z)\left( B^{-2} |(\nabla \times \mathbf{B}) \times \mathbf{B}|^2+
            |\nabla \cdot \mathbf{B}|^2\right)~d^3x,%
      \label{equ:optim}
      \end{equation}
where $V$ denotes the volume of the computational box and $w(x,y,z)$ is a weighting function
\citep[for details, see][]{wie_04}. The SOLIS data were preprocessed \citep[for details, see][]{wie_inh_06}
so that the forced photospheric boundary became closer to a force-free state to provide suitable boundary
conditions for the minimization of Eq.~(\ref{equ:optim}). For the corresponding potential and force-free
magnetic field, we can then estimate an upper limit to the free magnetic energy associated with coronal
currents of the form
      \begin{equation}
      E_{free} =\frac{1}{2\mu_0} \int_V\left( B_{nlff}^2 - B_{pot}^2\right)~ d^3x,%
      \label{equ:efree}
      \end{equation}
where $\mu_0$ denotes the magnetic permeability of vacuum, and $B_{pot}$ and $B_{nlff}$ represent the
total energy content of the potential and NLFF magnetic field, respectively. To estimate the uncertainty in the
numerical result, the code was applied to the original SOLIS data to which random, artificial noise had been
added in the form of a normal distribution of amplitude approximately equal to 1~G in the longitudinal and
50~G in the transversal component. The chosen noise amplitudes relate to the sensitivity of the VSM instrument.
It measures the Stokes V parameter far more accurately than the parameters Q and U. While the longitudinal
field is proportional to V, the transverse component is derived from Q and U, which are the principal source of
uncertainty.
\section{Results}
\subsection{Flare activity of NOAA AR 10960}
\begin{figure}[t]\centering
    \includegraphics[width=0.45\textwidth]{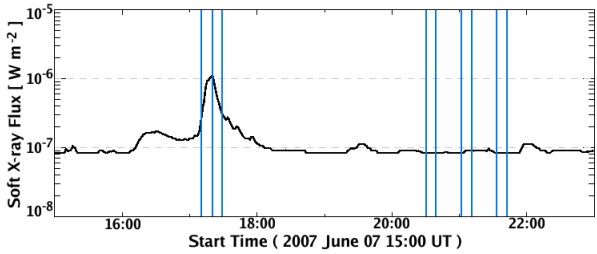}
    \caption{Solar SXR flux on 2007 June 7 in the wavelength range of 0.1 -- 0.8 nm. Vertical lines indicate
            the availability of SOLIS/VSM data.}
    \label{fig:fig2}
    \end{figure}

The solar activity during the week of 2007 June 4 was dominated by NOAA AR 10960. An M8.9 flare occurred
on June~4 and an M1.0 flare fired off on June 9. Furthermore, 12 C-class flares were detected during this week
and originated in this group or from its vicinity. The peaks in the measured solar soft X-ray (SXR) flux indicated
only one C1.0 flare on June~7, peaking at 17:20~UT, which is from interest for the present study (see
Fig.~\ref{fig:fig2}). SOLIS data with a high time cadence are available only for June 7.
\subsection{Global magnetic energy budget}
\begin{figure}[t]\centering
        \centerline{\includegraphics[width=0.45\textwidth]{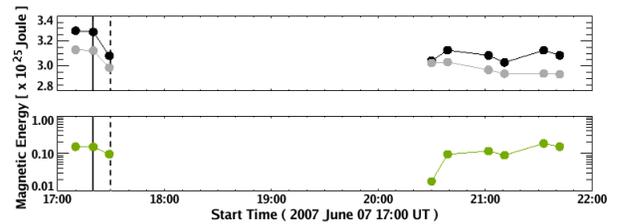}}
        \caption{\textit{Upper panel:} magnetic energy of the potential (gray) and NLFF (black) field.
        \textit{Lower panel:} upper limit for the free magnetic energy (shown on logarithmic scale).
        Solid and dashed lines represent the recorded C1.0 flare and CME, respectively.}
        \label{fig:fig3}
    \end{figure}

\begin{figure}[ht]\centering
        \centerline{\includegraphics[width=0.3\textwidth]{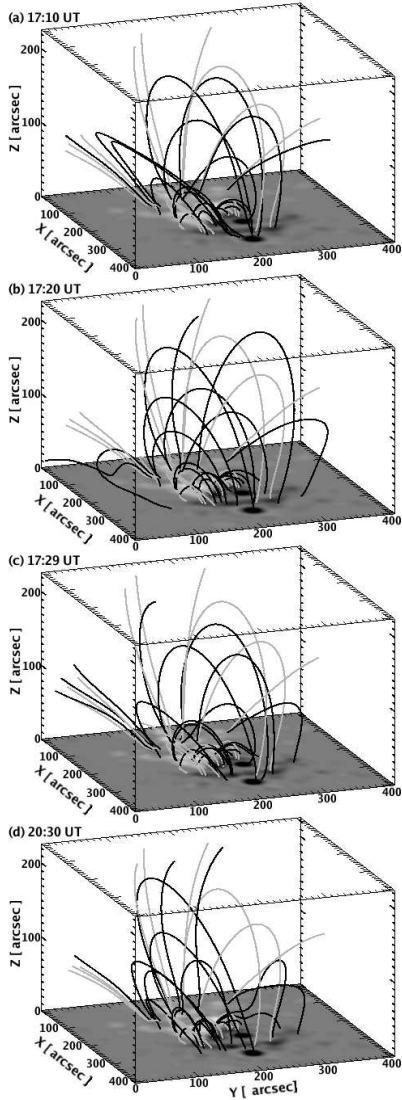}}
        \caption{Panels (a), (b), and (c) show the magnetic field configuration during the C1.0 flare. Panel (d)
        shows the minimum energy configuration. Shown are field lines of the potential (gray) and NLFF
        (black) field. For improved visibility, the $z$-axis is drawn elongated.}
        \label{fig:fig4}
    \end{figure}

For all magnetic field configurations, we find  that the energy of the extrapolated NLFF field exceeds that of the
potential field (i.e., $E_{nlff} > E_{pot}$), both being approximately $10^{25}$~J (see Table~\ref{tab:tab1}).
This is also the case when considering the evaluated relative error in the energy estimation of about 0.4 \% for the
potential and 1 \% for the NLFF field (i.e. $E_{pot}~\pm$ 0.013 $\times~10^{25}$~J and
$E_{nlff}~\pm$ 0.032 $\times~10^{25}$~J, respectively). The available free magnetic energy is always
approximately $10^{24}$~J with a relative error of about 14~\% (i.e. $E_{free}~\pm$ 0.026
$\times~10^{24}$~J). These uncertainty ranges were checked by comparing similar
results for 3D fields during the quiet period, and once by calculating the energy variation in the force-free fields
after adding artificial noise to the original magnetograms. Both $E_{pot}$ and $E_{nlff}$ were highest in the
phase of increasing emission from  the C1.0 flare. During the 20 minute time period of the flare
(17:10~UT -- 17:29~UT, see Fig.~\ref{fig:fig3}) the magnetic energy decreased by
$\Delta E_{nlff} = 2.01 \times 10^{24}$~J, i.e. $\approx$~38~\%  of the available free magnetic energy was
released. Although the  flare was already declining in intensity at 17:29~UT, it still showed a SXR flux of above
background B-level (see Fig.~\ref{fig:fig2}). The next vector magnetogram snapshot was acquired only 3 hours
later at 20:30~UT (see panel (d) of Fig.~\ref{fig:fig1}) and the free magnetic energy had decreased further, such
that $\approx$ 88.16~\% of the original amount of free energy had been released. At 20:30~UT, the magnetic
energy was only $\approx$ 0.6~\% of the total energy and consequently the magnetic field was almost
potential.\\
From Fig.~\ref{fig:fig2}, we can see that AR10960 showed only background B-level activity (i.e. a SXR emission
$<$ 10$^{-7}$ Wm$^{-2}$) at about 18:15~UT and almost the entire free magnetic energy may have been
released by that time. Unfortunately,  no vector magnetograph data was available immediately after the declining
phase of the flare and we are unable to confirm this supposition. For the AR studied here, the maximum excess
energy of a NLFF field over the potential field was  about 5~\% during the investigated period. Since this so-called
free energy is an upper limit to the available energy to drive eruptive phenomena, consequently only a small C1.0
flare was recorded. No further flares occurred between 20:30~UT and 21:42~UT for which SOLIS data is
available, but five C-class flares were recorded about 3 hours later on 2007 June 8 between 01:00~UT and
16:00~UT. However, a significant amount of free magnetic energy accumulated again during the quiet period
after 20:30~UT (see Table~\ref{tab:tab1} and Fig.~\ref{fig:fig3}) so that the energy content of the field
increased and became, with some fluctuation, comparable to that measured before the C1.0 flare.\\
From a visual inspection of the magnetic field lines within the extrapolation volume, we recognize some changes
in the magnetic field structure during the C1.0 flare (see Fig.~\ref{fig:fig4}). We find that the field lines show
their highest vertical extent when the C1.0 flare peaked (panel (b)). Ten minutes earlier a comparable field
structure was found (panel (a)), but with a lower vertical extend. Nine minutes after the flare peaked, the field
line configuration reached its lowest vertical extent (panel (c)), and more field lines of the NLFF field left the
extrapolation volume. However, at 20:30~UT, which corresponds to the field configuration of the lowest energy
content, the field clearly appears to have restructured, reaching on average its lowest altitude. The return to an
almost potential structure could be assigned to a coronal mass ejection (CME) recorded by the SoHO/LASCO
instrument on 2007~June~7 around 17:30~UT which could have bodily removed magnetic helicity of the coronal
field.
    \begin{table}[t]
    \caption{Magnetic energy of the extrapolated field.}\label{tab:tab1}\centering
    \vspace{-0.2cm}
    \begin{tabular}{ccccc}
        \hline
        \noalign{\smallskip}
        Time & \multicolumn{3}{c}{~~~~~Magnetic energy $\mathrm{[~\times~10^{25}~J~]}$}&~\\
        $\mathrm{(~UT~)}$&~~~~~$E_{pot}^{(1)}$ &~~~~~~$E_{nlff}^{(2)}$%
        & ~~~~~$E_{free}^{(3)}$ & ~~~$E_{nlff}/E_{pot}^{(4)}$ \\
        \noalign{\smallskip}
        \hline
        \noalign{\smallskip}
        17:10       & ~~~~~3.130      & ~~~~~3.282     & ~~~~~ 0.152     & ~~ 1.049\\ %
        17:20       & ~~~~~3.122      & ~~~~~3.272     & ~~~~~ 0.149     & ~~ 1.048\\ %
        17:29       & ~~~~~2.986      & ~~~~~3.081     & ~~~~~ 0.095     & ~~ 1.032 \\ %
        20:30       & ~~~~~3.024      & ~~~~~3.042     & ~~~~~ 0.018     & ~~ 1.006\\ %
        20:39       & ~~~~~3.031      & ~~~~~3.127     & ~~~~~ 0.095     & ~~ 1.031\\ %
        21:02       & ~~~~~2.969      & ~~~~~3.084     & ~~~~~ 0.116     & ~~1.039\\ %
        21:11       & ~~~~~2.938      & ~~~~~3.028     & ~~~~~ 0.090     & ~~1.031\\ %
        21:33       & ~~~~~2.939      & ~~~~~3.125     & ~~~~~ 0.185     & ~~1.063\\ %
        21:42       & ~~~~~2.933      & ~~~~~3.085     & ~~~~~ 0.152     & ~~1.052\\ %
        \noalign{\smallskip}
        \hline
    \end{tabular}
    \par \smallskip \footnotesize{$^{(1)}$ Energy of the potential and $^{(2)}$ NLFF field, $^{(3)}$ upper
        limit of the free energy, and $^{(4)}$ excess energy of the NLFF over the potential field.}
    \end{table}
\section{Discussion}
We have investigated the coronal magnetic field associated with the NOAA AR 10960 on 2007 June 7 by
analyzing SOLIS/VSM data. Three vector magnetograms with a time cadence of  $\approx$ 10 minutes were
available to investigate the magnetic energy content of the coronal field during a C1.0 flare, and six further
snapshots were acquired to analyze a very quiet time about three hours after the flare. Before as well as after the
small flare, the magnetic field energy was $E_{nlff} \approx 3 \times 10^{25}$~J. The NLFF field had a free
energy of $E_{free} \approx 1.5 \times 10^{24}$ J before the flare. As a consequence of the flare/CME, this free
magnetic energy reduced by almost a factor of 10 and produced an almost potential configuration. Six snapshots
acquired within a time period of about $70$ minutes, during a quiet period of 3 -- 4 hours after the flare, showed
again an increase in the free magnetic energy. Since the estimated free magnetic energy remained only about
5~\% of the total energy content, no large eruption was produced by AR 10960.\\
This is clearly different from the flaring of AR 10540 observed on 2004 January 18 -- 21, which was analyzed in a
previous work with the help of vector magnetograph data from the Solar Flare Telescope in Japan of time
cadence of about 1 day. In this AR, the free energy was $E_{free}\approx$ 66~\% of the total energy, which was
sufficiently high to power a M6.1 flare \citep[for details see][]{tha_08}. The activity of AR 10540 investigated
earlier was significantly higher than for the data analyzed in the current paper, as was the total magnetic energy.
However, despite these differences, we also found some common features. Magnetic energy accumulates before the
flare and a significant part of the excess energy is released during the flare. The high amount of free magnetic
energy available in AR 10540 produced M-class flares, while the relatively small amount of free energy in AR
10960 powered only a small C-class flare. In both cases, all three components of the vector magnetogram changed
during the flare, but the energy decrease in the NLFF field was always higher than that of the potential field, i.e.
the energy release was more related to the change in the transverse magnetic field components -- which
correspond to the field aligned electric currents in the corona -- than to that of the longitudinal component.
\begin{acknowledgements}
\footnotesize
SOLIS/VSM vector magnetograms are produced cooperatively by NSF/NSO and NASA/LWS. The National Solar
Observatory (NSO) is operated by the Association of Universities for Research in Astronomy, Inc., under cooperative
agreement with the National Science Foundation. J.~K.~Thalmann is supported by DFG-grant WI 3211/1-1,
T.~Wiegelmann by DLR-grant 50 OC 0501, and N.-E.~Raouafi by NSO and NASA grant NNH05AA12I. We thank
B.~Inhester for helpful discussions.
\end{acknowledgements}
\bibliographystyle{aa}

{\small
\begin{thebibliography}{18}
\expandafter\ifx\csname natexlab\endcsname\relax\def\natexlab#1{#1}\fi

\bibitem[{{Alissandrakis}(1981)}]{ali_81}
{Alissandrakis}, C.~E. 1981, \aap, 100, 197

\bibitem[{{Amari} {et~al.}(1997){Amari}, {Aly}, {Luciani}, {Boulmezaoud}, \&
  {Mikic}}]{ama_97}
{Amari}, T., {Aly}, J.~J., {Luciani}, J.~F., {Boulmezaoud}, T.~Z., \& {Mikic},
  Z. 1997, \solphys, 174, 129

\bibitem[{{Georgoulis}(2005)}]{geo_05}
{Georgoulis}, M.~K. 2005, \apjl, 629, L69

\bibitem[{{Georgoulis} {et~al.}(2008){Georgoulis}, {Raouafi}, \&
  {Henney}}]{geo_08}
{Georgoulis}, M.~K., {Raouafi}, N.-E., \& {Henney}, C.~J. 2008, in ASP Conf.
  Ser., ed. R.~{Howe}, R.~W. {Komm}, K.~S. {Balasubramaniam}, \& G.~J.~D.
  {Petrie}, Vol. 383, 107

\bibitem[{{Henney} {et~al.}(2006){Henney}, {Keller}, \& {Harvey}}]{hen_06}
{Henney}, C.~J., {Keller}, C.~U., \& {Harvey}, J.~W. 2006, in ASP Conf. Ser.,
  ed. R.~{Casini} \& B.~W. {Lites}, Vol. 358, 92

\bibitem[{{Jones} {et~al.}(2002){Jones}, {Harvey}, {Henney}, {Hill}, \&
  {Keller}}]{jon_02}
{Jones}, H.~P., {Harvey}, J.~W., {Henney}, C.~J., {Hill}, F., \& {Keller},
  C.~U. 2002, in ESA SP, Vol. 505, Magnetic Coupling of the Solar Atmosphere,
  ed. H.~{Sawaya-Lacoste}, 15

\bibitem[{{Keller} {et~al.}(2003){Keller}, {Harvey}, \&
  {Giampapa}}]{kel_har_03}
{Keller}, C.~U., {Harvey}, J.~W., \& {Giampapa}, M.~S. 2003, in Innovative
  Telescopes and Instrumentation for Solar Astrophysics., ed. S.~L. {Keil} \&
  S.~V. {Avakyan}, Vol. 4853, 194

\bibitem[{{Metcalf} {et~al.}(2008){Metcalf}, {Derosa}, {Schrijver}, {Barnes},
  {van Ballegooijen}, {Wiegelmann}, {Wheatland}, {Valori}, \&
  {McTtiernan}}]{met_08}
{Metcalf}, T.~R., {Derosa}, M.~L., {Schrijver}, C.~J., {et~al.} 2008, \solphys,
  247, 269

\bibitem[{{R{\'e}gnier} \& {Priest}(2007)}]{reg_07b}
{R{\'e}gnier}, S. \& {Priest}, E.~R. 2007, \apjl, 669, L53

\bibitem[{{Sakurai}(1989)}]{sak_89}
{Sakurai}, T. 1989, Space Science Reviews, 51, 11

\bibitem[{{Schrijver} {et~al.}(2008){Schrijver}, {DeRosa}, {Metcalf}, {Barnes},
  {Lites}, {Tarbell}, {McTiernan}, {Valori}, {Wiegelmann}, {Wheatland},
  {Amari}, {Aulanier}, {Demoulin}, {Fuhrmann}, {Kusano}, {Regnier}, \&
  {Thalmann}}]{sch_08}
{Schrijver}, C.~J., {DeRosa}, M.~L., {Metcalf}, T., {et~al.} 2008, \apj, 675,
  1637

\bibitem[{{Schrijver} {et~al.}(2006){Schrijver}, {Derosa}, {Metcalf}, {Liu},
  {McTiernan}, {R{\'e}gnier}, {Valori}, {Wheatland}, \& {Wiegelmann}}]{sch_06}
{Schrijver}, C.~J., {Derosa}, M.~L., {Metcalf}, T.~R., {et~al.} 2006, \solphys,
  235, 161

\bibitem[{{Thalmann} \& {Wiegelmann}(2008)}]{tha_08}
{Thalmann}, J.~K. \& {Wiegelmann}, T. 2008, \aap, 484, 495

\bibitem[{{Wheatland} {et~al.}(2000){Wheatland}, {Sturrock}, \&
  {Roumeliotis}}]{whe_00}
{Wheatland}, M.~S., {Sturrock}, P.~A., \& {Roumeliotis}, G. 2000, \apj, 540,
  1150

\bibitem[{{Wiegelmann}(2004)}]{wie_04}
{Wiegelmann}, T. 2004, \solphys, 219, 87

\bibitem[{{Wiegelmann}(2008)}]{wie_08}
{Wiegelmann}, T. 2008, JGR (Space Physics), 113, 3

\bibitem[{{Wiegelmann} {et~al.}(2006){Wiegelmann}, {Inhester}, \&
  {Sakurai}}]{wie_inh_06}
{Wiegelmann}, T., {Inhester}, B., \& {Sakurai}, T. 2006, \solphys, 233, 215

\bibitem[{{Wiegelmann} {et~al.}(2005){Wiegelmann}, {Lagg}, {Solanki},
  {Inhester}, \& {Woch}}]{wie_05}
{Wiegelmann}, T., {Lagg}, A., {Solanki}, S.~K., {Inhester}, B., \& {Woch}, J.
  2005, \aap, 433, 701

\end{thebibliography}
}

%
\end{document}